\documentclass[iop]{emulateapj}

\journalinfo{ApJ in press}
\submitted{Received 2013 March 8; accepted 2013 May 31}

\shorttitle{EVIDENCE FOR TIDAL DESTRUCTION}
\shortauthors{SCHLAUFMAN \& WINN}

\begin{document}

\title{EVIDENCE FOR THE TIDAL DESTRUCTION OF HOT JUPITERS BY SUBGIANT STARS}

\author{Kevin C.\ Schlaufman\altaffilmark{1} and Joshua N.\ Winn\altaffilmark{2}}
\affil{Kavli Institute for Astrophysics and Space Research,
Massachusetts Institute of Technology, Cambridge, Massachusetts 02139,
USA}
\email{kschlauf@mit.edu, jwinn@mit.edu}

\altaffiltext{1}{Kavli Fellow}
\altaffiltext{2}{Physics Department, Massachusetts Institute of Technology,
Cambridge, Massachusetts 02139, USA}

\begin{abstract}

\noindent
Tidal transfer of angular momentum is expected to cause hot Jupiters
to spiral into their host stars.  Although the timescale for orbital
decay is very uncertain, it should be faster for systems with larger
and more evolved stars.  Indeed, it is well established that hot
Jupiters are found less frequently around subgiant stars than around
main-sequence stars.  However, the interpretation of this finding has
been ambiguous, because the subgiants are also thought to be more massive
than the F- and G-type stars that dominate the main-sequence sample.
Consequently it has been unclear whether the absence of hot Jupiters
is due to tidal destruction, or inhibited formation of those planets
around massive stars.  Here we show that the Galactic space motions
of the planet-hosting subgiant stars demand that on average they be
similar in mass to the planet-hosting main-sequence F- and G-type stars.
Therefore the two samples are likely to differ only in age, and provide
a glimpse of the same exoplanet population both before and after tidal
evolution.  As a result, the lack of hot Jupiters orbiting subgiants is
clear evidence for their tidal destruction.  Questions remain, though,
about the interpretation of other reported differences between the
planet populations around subgiants and main-sequence stars, such as
their period and eccentricity distributions and overall occurrence rates.

\end{abstract}


\section{Introduction}

Although radial velocity planet surveys have mainly targeted main-sequence
FGKM stars, there have also been numerous discoveries of Jupiter-mass
giant planets around evolved stars.  These include moderately evolved
subgiant stars, as well as very evolved giant stars.\footnote{For subgiant
stars, see for example \citet{hat03}, \citet{joh07}, \citet{rob07},
  \citet{fis07}, \citet{joh08}, \citet{bow10}, \citet{joh10a,joh10c}, and
  \citet{joh11a,joh11b}.  For giant stars, see for example \citet{hat93},
  \citet{fri02}, \citet{sat03}, \citet{set05}, \citet{but06}, \citet{hat06},
  \citet{dol07}, \citet{sat07}, \citet{liu08}, \citet{sat08}, \citet{dol09},
  \citet{liu09}, \citet{han10}, \citet{sat10}, \citet{lee11,lee13}, and
  \citet{omi12}.}  This work has provided evidence that the giant planet
population around evolved stars differs from the population orbiting
solar-type main-sequence stars, in at least three respects.  First,
there are fewer close-in giant planets (``hot Jupiters'') around evolved
stars than main-sequence stars \citep[e.g.,][]{bow10,joh10a}.  Second,
the orbital eccentricities of long-period giant planets are typically
lower when the host star is evolved.  Third, a larger fraction of evolved
stars appear to host long-period giant planets, although this comparison
is more complicated because of the possibility of systematic metallicity
differences between the samples of evolved stars and main-sequence stars
\citep[e.g.,][]{joh10b}.

The work described here was motivated by the desire to understand the
first finding, the scarcity of hot Jupiters around evolved stars. Two
conflicting interpretations have been proposed.  The first possibility
is that tidal evolution has destroyed the hot Jupiters that once orbited
the evolved stars \citep[e.g.,][]{ras96,vil09,kun11,ada12}.  The second
possibility is that the evolved stars are on average more massive than the
main-sequence stars, and the differences in the planet populations are
linked to the enhanced stellar mass \citep[e.g.,][]{bur07,kre09,cur09}.
In this picture, hot Jupiters may occur less commonly around massive stars
due to differences in (for example) the structure of the protoplanetary
disk, or the disk dissipation timescale.

The latter scenario, attributing the differences to stellar mass, is
generally favored in the literature.  It is supported by the results of
fitting stellar-evolutionary models to the observable properties of the
evolved stars (their luminosities, surface gravities, and effective
temperatures), which confirm that the evolved stars are relatively
massive.  \cite{joh07} referred to their sample of evolved stars as
``retired A stars,'' because the evolutionary models assign these stars
the same masses as main-sequence A5--A9 stars (2.0--1.6~$M_\odot$).
Recently, though, the fidelity of those evolutionary models was called
into question by \citet{llo11}.  He argued that the selection criteria
that were used to define samples of evolved stars should have resulted
in a sample dominated by lower-mass stars, with F- and G-type progenitors
rather than A-type.  The debate over this claim continues \citep{joh13}.
In the meantime, it would be helpful to have a model-independent method
for comparing the masses of the evolved stars and the main-sequence
stars that have been included in radial velocity surveys.

The observed Galactic space motions (kinematics) of the stars can
provide such a comparison.  All of the stars under discussion are
members of the thin disk of the Milky Way, and empirically it is known
that the velocity dispersion of a thin-disk population increases with
age \citep[e.g.,][]{bin00}.  This is understood as follows.  Thin disk
stars form from dense, turbulent gas in the Galactic plane.  Because that
process is highly dissipative, stellar populations are formed with a
very cold velocity distribution.  Over time, the velocity distribution
is heated due to interactions between the stars and molecular clouds
\citep[e.g.,][]{spi51} and transient spiral waves \citep[e.g.,][]{bar67}.
Massive stars spend only a short time on the main sequence, and there
is very little time for collisions to kinematically heat a population of
massive stars.  On the other hand, solar-mass stars spend a long time on
the main sequence, leaving plenty of time for collisions to kinematically
heat a population of solar-mass stars.  One would therefore expect a
main-sequence thin-disk stellar population's space velocity dispersion
to decrease with increasing stellar mass.  The same is true even for
evolved stars, because a star spends such a small fraction of its life
as a subgiant or giant relative to its main-sequence lifetime.

In this paper, we investigate the Galactic velocity dispersion of the
population of evolved planet-hosting stars.  The kinematic evidence shows
that these stars are similar in mass to F5--G5 main-sequence stars.
Consequently, the lack of host Jupiters around evolved stars cannot
be attributed to mass.  The most plausible explanation is that the hot
Jupiters have been destroyed, after losing orbital angular momentum to
their host stars.  We describe our sample of planet-hosting stars in
Section 2, we detail our analysis procedures in Section 3, we outline
scaling relations for tidal evolution timescales in Section 4, we discuss
the results and implications in Section 5, and we summarize our findings
in Section 6.

\section{Sample definition}

We first extract a list of exoplanet host stars identified with the
Doppler technique, using the online database \texttt{exoplanets.org}
\citep{wri11}.  We cross-match those stars with the Hipparcos
catalog \citep{van07} to obtain parallaxes and $B-V$ colors, and
with the Tycho-2 catalog \citep{hog00} to obtain apparent $V$-band
magnitudes.  We transform the Tycho-2 $B_{T}$ and $V_{T}$ magnitudes
into approximate Johnson-Cousins $V$-band magnitudes using the relation
$V = V_{T} - 0.090\left(B_{T}-V_{T}\right)$.  Photometry for some of
the brightest planet-hosting stars is missing from the \citet{hog00}
catalog.  For those cases we use the original Tycho photometry given
by \citet{per97}.  We remove from the list all planet-hosting stars with
imprecise parallaxes (fractional uncertainty exceeding 20\%).  We also
obtain the nominal parameters of the host stars and their planets from
\texttt{exoplanets.org}.

\begin{figure*}
\plotone{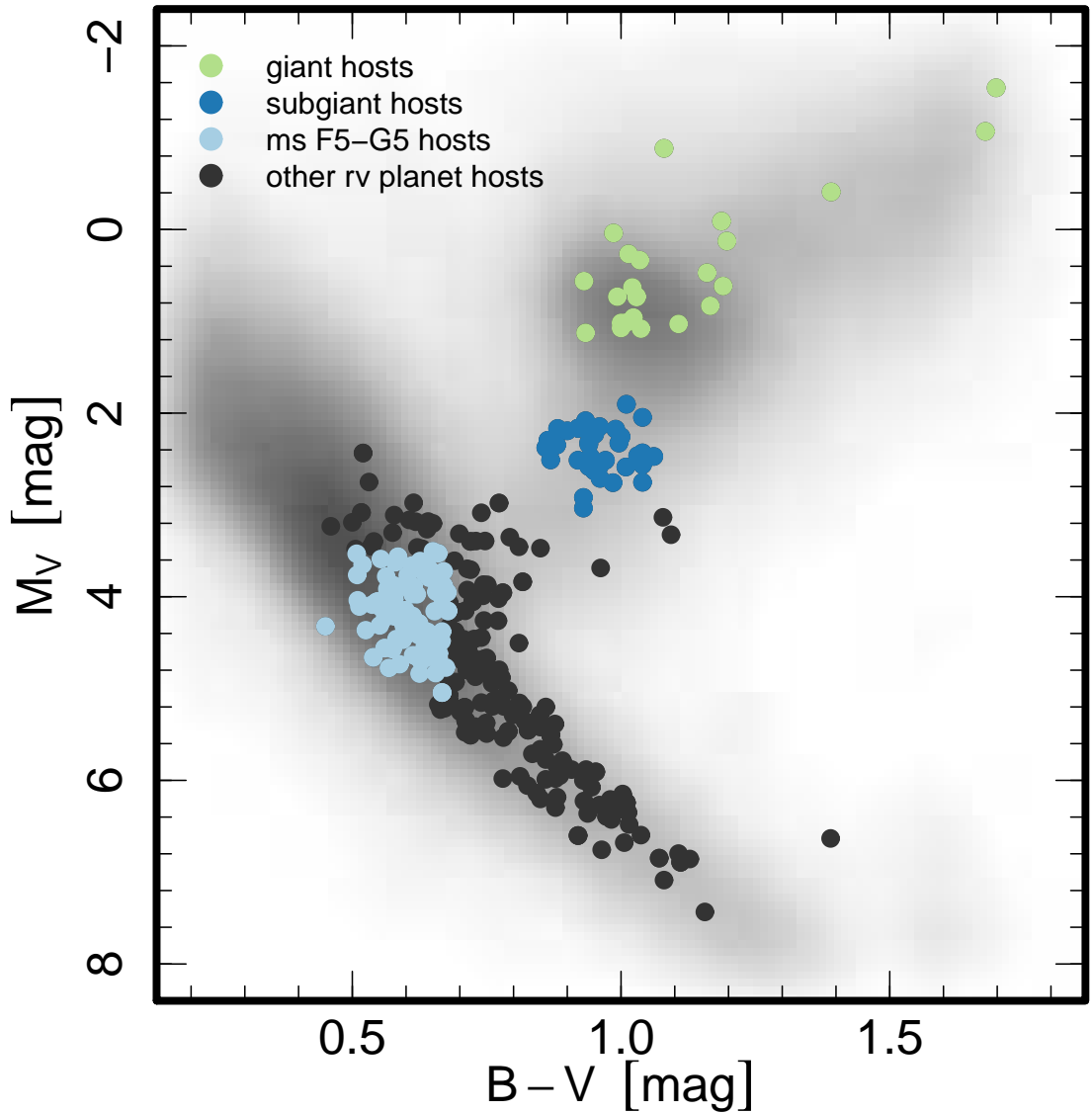}
\caption{Hertzsprung-Russell diagram of exoplanet host stars identified
with the radial-velocity technique listed at \texttt{exoplanets.org}
\citep{wri11}.  We isolate and plot three subsamples of the full exoplanet
host population: F5--G5 main-sequence stars (light blue points),
subgiant stars (dark blue points), and giant stars (green points).
We plot all other exoplanet host stars as black points.  We plot only
those stars that have Hipparcos parallaxes with an uncertainty smaller
than 20\%.  The background shading indicates the density of stars in the
entire Hipparcos catalog with relative parallax uncertainties smaller
than 20\%.  The $B-V$ colors come from \citet{van07} and the $V$-band
photometry is derived from Tycho-2 magnitudes given in \citet{hog00}
according to the relation $V = V_{T} - 0.090\left(B_{T}-V_{T}\right)$.
Following \citet{bin98}, we define the F5--G5 sample as those stars
with $0.44<\left(B-V\right)<0.68$ and $3.5<M_{V}<5.1$.   We define
the subgiant sample as those stars with $0.85<\left(B-V\right)<1.1$
and $1.6<M_{V}<3.1$.  We define the giant sample as those stars with
$M_{V}<1.6$.\label{fig01}}
\end{figure*}

We then define three samples of the planet-hosting stars.  The sample
of subgiants is defined as those stars with
$0.85<\left(B-V\right)<1.1$ and $1.6<M_{V}<3.1$.  The sample of giants
consists of stars with $M_{V}<1.6$.  Finally, a sample of
main-sequence F5--G5 planet-hosting stars is defined by the criteria
$0.44<\left(B-V\right)<0.68$ and $3.5<M_{V}<5.1$, taken from
\citet{bin98}. We will show later, based on kinematic evidence, that
the typical mass of stars in the main-sequence F5--G5 sample is
similar to the typical mass of the stars in the subgiant and giant
star samples.\footnote{We also constructed samples corresponding to
  F0--F5, F5--G0, and G0--G5 stars. We settled on F5--G5 for the
  analysis presented here, because this sample gave the best match to
  the kinematics of the subgiant planet hosts; see Section 3.1.}
Figure~\ref{fig01} is a Hertzsprung-Russell (HR) diagram of
planet-hosting stars identified with the radial velocity technique,
with our three subsamples highlighted.

Our samples are defined according to observable criteria in color and
absolute magnitude, as opposed to model-dependent stellar masses
(i.e., masses determined by comparing the observable properties with
the outputs of theoretical stellar-evolutionary models). This is
because the planet surveyors often define samples according to color
and absolute magnitude, and because of the questionable reliability of
the model-dependent stellar masses for the subgiants (Lloyd~2011).
Nevertheless, one may wonder about the model-dependent masses that
have been determined for the stars in our samples.  For the subgiants,
the model-dependent masses reported in the literature are in the range
$0.92$--$2.0~M_{\odot}$, with all but six stars having
$M_{\ast}>1.4~M_{\odot}$.\footnote{If we remove the six stars with
  model-dependent masses smaller than 1.4~$M_\odot$ from this sample,
  none of the subsequent analyses are substantially affected.} The
giant stars have model-dependent masses reported to be in the range
$0.92$--$2.7~M_{\odot}$, with all but 6 stars having
$M_{\ast}>1.4~M_{\odot}$.  The reported model-dependent masses of the
main-sequence F5--G5 stars are in the range $0.8$--$1.4~M_{\odot}$.

We also define two samples of main-sequence stars in the solar
neighborhood (not necessarily planet-hosting stars) using the Hipparcos
catalog, requiring as before that the parallax uncertainty be smaller
than 20\%. The first sample consists of main-sequence stars with
$0.15<(B-V)<0.30$ and $1.9<M_{V}<2.7$, corresponding to A5--F0 stars
(2--1.5 $M_{\odot}$) according to \citet{bin98}. Our intention with
this sample is to provide a control sample that will allow a test of the
hypothesis that the population of subgiant planet hosts is dominated by
``retired A stars''. The second sample consists of main-sequence stars
with $0.44<(B-V)<0.68$ and $3.5<M_{V}<5.1$, corresponding to F5--G5 stars
(1.3--0.9 $M_{\odot}$) according to \citet{bin98}.

We compute Galactic $UVW$ velocities for all stars for which the systemic
radial velocities are available in the catalogs of \citet{gon06},
\citet{mas08}, or \citet{chu12}. In addition to radial velocities,
as inputs we use Hipparcos astrometry, parallaxes, and proper motions
\citep{van07}. Enough information is available to compute $UVW$ velocities
for all 35 subgiants, 23 of 24 giant stars, and 88 of 91 main-sequence
F5--G5 stars. We use the algorithm given in \texttt{gal\_uvw.pro}
available from The IDL Astronomy User's Library.  We follow the convention
that $U$ is positive towards the Galactic center, $V$ is positive in
the direction of Galactic rotation, and $W$ is positive towards the
North Galactic pole.  We do not correct for motion of the Sun relative
to the local standard of rest.  For those interested in replicating
the results, a comparable data set is publicly available from E.\
Mamajek\footnote{www.pas.rochester.edu/\textasciitilde
  emamajek/HIP2008\_UVW\_SpT\_Mv.dat}.

\section{Sample comparisons}

\subsection{Kinematics}

\begin{figure*}
\plotone{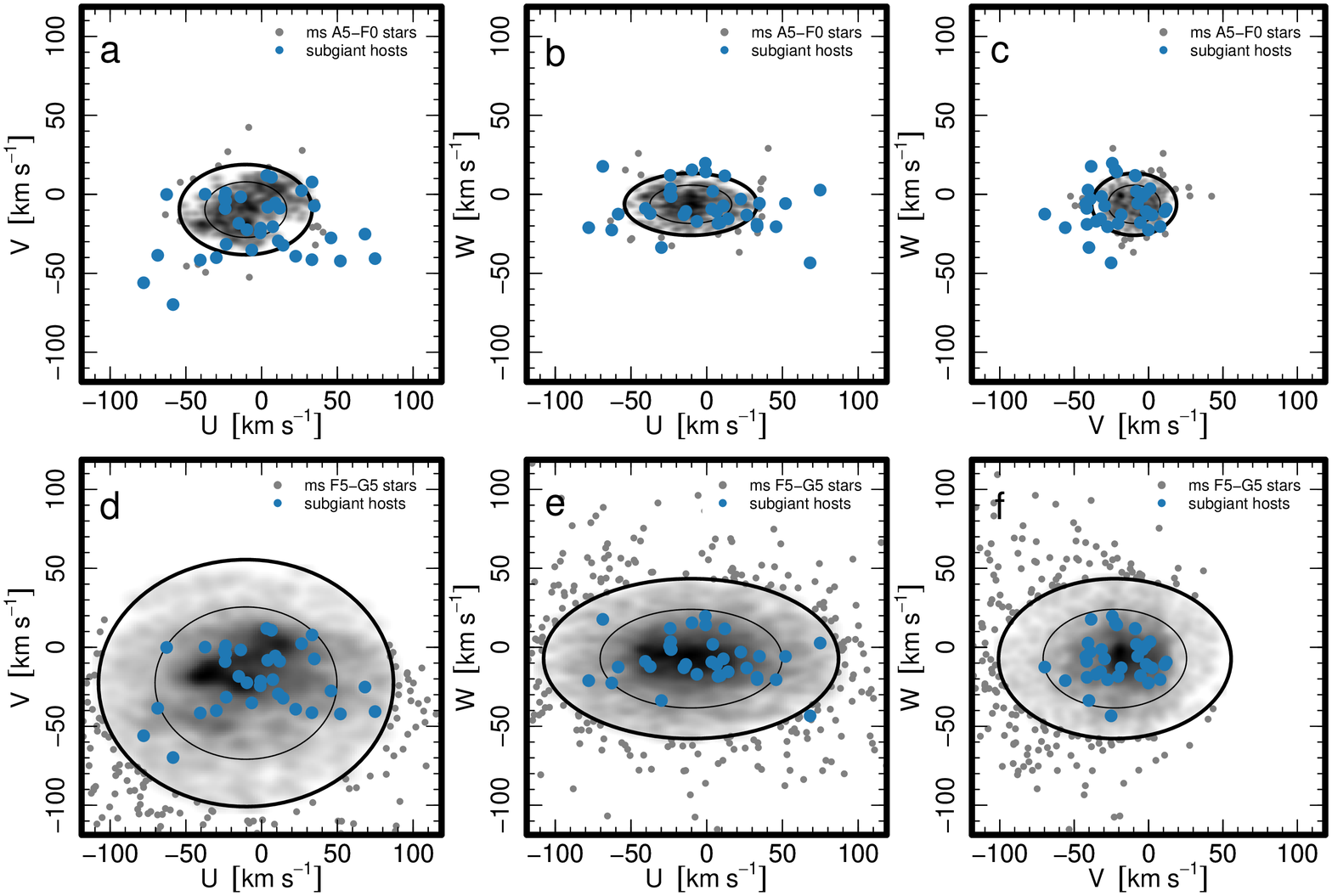}
\caption{Galactic $UVW$ kinematics of subgiant stars that host
exoplanets discovered with the radial-velocity technique.  In each
panel, we plot the $UVW$ space motions of the subgiant sample as blue
points and the density of points in a control sample selected from the
Hipparcos catalog as the background shading.  We plot outliers in the
control sample as gray points.  In panels a, b, and c the control sample
consists of main-sequence A5--F0 Hipparcos stars with $0.15<(B-V)<0.30$
and $1.9<M_{V}<2.7$ \citep[e.g.,][]{bin98}.  In panels d, e, and f the
control sample consists of main-sequence Hipparcos F5--G5 stars with
$0.44<(B-V)<0.68$ and $3.5<M_{V}<5.1$ \citep[e.g.,][]{bin98}.  In each
panel, the light (heavy) ellipse is the 68\% (95\%) velocity ellipsoid
of the control sample in that panel.  The $UVW$ velocity distribution
of the subgiant sample is kinematically much hotter than the $UVW$
velocity distribution of the main-sequence A5--F0 sample.  However,
it is well matched by the $UVW$ velocity distribution of the main
sequence F5--G5 stars.  We quantify these observations in Section 3.1.
The velocity dispersion of a thin disk stellar population increases
with age \citep[e.g.,][]{bin00}.  Since the post--main-sequence phase
of stellar evolution is much shorter than the main-sequence phase, the
velocity dispersion of a post--main-sequence stellar population depends
on the main-sequence lifetimes of its constituent stars.  As a result,
the large velocity dispersion of subgiant exoplanet host sample relative
to the main-sequence A5--F0 population indicates that subgiants spent
more time on the main sequence than the A5--F0 stars.  Therefore the
subgiants are on average lower in mass than the A5--F0 sample, and likely
to be similar in mass to the F5--G5 sample.\label{fig02}}
\end{figure*}

Figure~\ref{fig02} compares the $UVW$ velocities of the subgiant
planet-hosting stars, the solar-neighborhood A5--F0 stars, and the
solar-neighborhood F5--G5 stars. As expected, the F5--G5 stars have
a much larger velocity dispersion than the A5--F0 stars; they are
kinematically hotter. The subgiant planet-hosting stars are also seen
to be kinematically hotter than the main-sequence A5--F0 stars, with
many stars outside of the 95\% ellipse defined by the A5--F0 population.
On the other hand, the velocity dispersion of the subgiant planet-hosting
stars is comparable to the velocity dispersion of the main-sequence
F5--G5 sample.

We quantify the significance of these observations in the following way.
First, we calculate the mean $UVW$ velocities of the solar neighborhood
main-sequence A5--F0 and F5--G5 samples. We denote these mean velocities
by $\bar{U}$, $\bar{V}$, and $\bar{W}$.  We then compute
\begin{eqnarray} \delta_{i} & = &
\left[\left(U_i-\bar{U}\right)^2+\left(V_i-\bar{V}\right)^2+\left(W_i-\bar{W}\right)^2\right]^{1/2}
\end{eqnarray}

\noindent
where $U_{i}$, $V_{i}$, and $W_{i}$ are the individual $UVW$ velocity
coordinates of a single star.  We calculate $\delta$ for each star
in the subgiant sample and $\delta'$ for each star in the solar
neighborhood A5--F0 sample.  We use the Anderson-Darling test \citep[see,
e.g.,][]{hou09} to compare the sample of subgiant $\delta_{i}$ values
to the sample of A5--F0 $\delta_{i}'$ values.  We find the probability
that the subgiant planet-hosting stars are drawn from the same parent
distribution as solar-neighborhood A5--F0 stars to be less than $10^{-6}$.
Then we repeat the same calculation after computing $\delta'$ for each
star in the solar neighborhood F5--G5 sample.  The result is that there
is no reason to reject the hypothesis that the subgiant planet hosts and
the F5--G5 stars are drawn from the same population (Anderson-Darling
$p$-value $=0.48$).

\begin{figure*}
\plotone{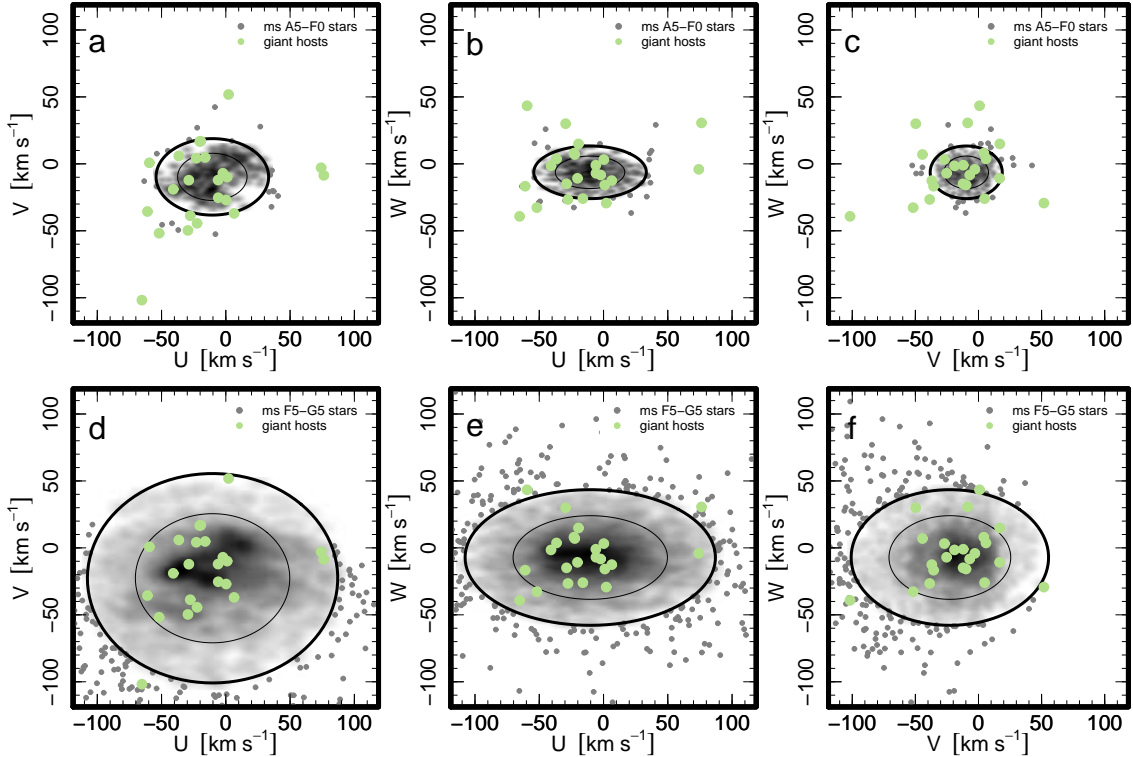}
\caption{Galactic $UVW$ kinematics of giant stars that host exoplanets
discovered with the radial-velocity technique.  In each panel, we plot the
$UVW$ space motions of the giant sample as green points and the density
of points in a control sample selected from the Hipparcos catalog as
the background shading.
We plot outliers in the control sample as gray points.  In panels a,
b, and c the control sample consists of main-sequence A5--F0 Hipparcos
stars with $0.15<(B-V)<0.30$ and $1.9<M_{V}<2.7$ \citep[e.g.,][]{bin98}.
In panels d, e, and f the control sample consists of main-sequence
Hipparcos F5--G5 stars with $0.44<(B-V)<0.68$ and $3.5<M_{V}<5.1$
\citep[e.g.,][]{bin98}.  In each panel, the light (heavy) ellipse is
the 68\% (95\%) velocity ellipsoid of the control sample in that panel.
The $UVW$ velocity distribution of the giant sample is kinematically much
hotter than the $UVW$ velocity distribution of the main-sequence A5--F0
sample.  However, it is well matched by the $UVW$ velocity distribution
of the main-sequence F5--G5 stars.  We quantify these observation in
Section 3.1.  As a result, the typical mass of a star in the giant
population is more likely to be similar in mass to a F5--G5 star than
an A5--F0 star.\label{fig03}}
\end{figure*}

The $UVW$ space motions of the subgiant planet-hosting stars are
incompatible with the hypothesis of recent evolution from a population
of A5--F0 main-sequence stars.  Their larger velocity dispersion
requires that they experienced longer main-sequence lifetimes than the
stars in the solar-neighborhood main-sequence A5--F0 sample.  This in
turn implies that the subgiant planet-hosting stars are less massive
than the main-sequence A5--F0 stars.  The agreement between the
velocity dispersions of the subgiants and the solar-neighborhood
main-sequence F5--G5 sample indicates that these two samples likely
have similar stellar masses.\footnote{When we use a control sample of
  F0--F5 stars rather than F5--G5 stars, the Anderson-Darling test
  indicates that the F0--F5 sample and the subgiant planet host sample
  have a probability $p=0.00033$ of being drawn from the same parent
  distribution. Thus the typical mass of the subgiant planet hosts is
  also likely to be smaller than that of the F0--F5 stars.}

This is at odds with the model-dependent masses for the subgiants, which
are generally reported to be larger than than the masses of main-sequence
F5--G5 stars.  While our sample-based test cannot be used to establish
the mass of any particular star, we note that there is only a very
weak negative correlation ($\approx\!\!-0.1$) between $\delta$ and the
model-dependent mass. This correlation is not significantly different
from zero given the sample size ($N=35$). The correlation would have
needed to be 0.3 or larger in absolute value to have established a
genuine correlation at the $p=0.05$ level.

Figure~\ref{fig03} compares the $UVW$ velocities of the giant
planet-hosting stars with the solar-neighborhood main-sequence stars.
The $UVW$ velocity dispersion of the giants is also inconsistent
with solar-neighborhood main-sequence A5--F0 sample (Anderson-Darling
$p<10^{-6}$).  And just as with the subgiants, their velocity dispersion
is indistinguishable from the solar-neighborhood main-sequence F5--G5
sample (Anderson-Darling $p=0.62$).  We conclude that the planet-hosting
giant stars, like the subgiants, are on average similar in mass to the
main-sequence F5--G5 planet hosts.

One of the planet-hosting giants, $\beta$ Geminorum (Pollux) happens
to have a asteroseismic mass determination placing it firmly in the
``retired A star'' category, with $M_{\ast} = 1.91\pm0.09~M_{\odot}$
\citep{hat12}.  Reassuringly, we find that this star has a $UVW$
velocity of $(-16,5,-26)$ km s$^{-1}$, placing it just outside of the
68\% contour of the solar-neighborhood main-sequence A5--F0 sample.
Therefore this particular star has a low space velocity as expected,
and does not call into question the overall result that the sample of
planet-hosting giants is dominated by lower-mass stars.

\begin{figure*}
\plotone{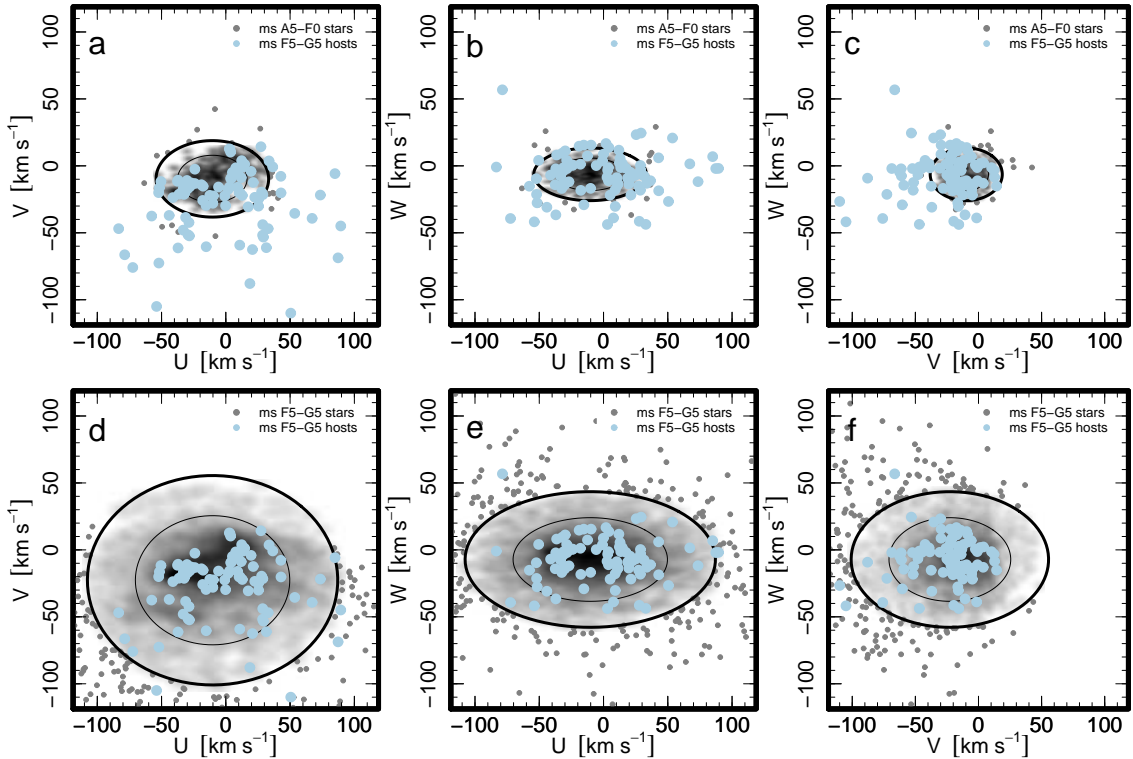}
\caption{Galactic $UVW$ kinematics of main-sequence F5--G5 stars that host
exoplanets discovered with the radial-velocity technique.  In each panel,
we plot the $UVW$ space motions of the main-sequence F5--G5 sample as
light blue points and the density of points in a control sample selected
from the Hipparcos catalog as the background shading.
We plot outliers in the control sample as gray points.  In panels a,
b, and c the control sample consists of main-sequence A5--F0 Hipparcos
stars with $0.15<(B-V)<0.30$ and $1.9<M_{V}<2.7$ \citep[e.g.,][]{bin98}.
In panels d, e, and f the control sample consists of main-sequence
Hipparcos F5--G5 stars with $0.44<(B-V)<0.68$ and $3.5<M_{V}<5.1$
\citep[e.g.,][]{bin98}.  In each panel, the light (heavy) ellipse is
the 68\% (95\%) velocity ellipsoid of the control sample in that panel.
As expected, the $UVW$ velocity distribution of the main-sequence F5--G5
sample is kinematically much hotter than the $UVW$ velocity distribution
of the main-sequence A5--F0 sample but well matched by the $UVW$ velocity
distribution of the larger Hipparcos main-sequence F5--G5 star sample.
We quantify these observations in Section 3.1.\label{fig04}}
\end{figure*}

Figure~\ref{fig04} compares the $UVW$ velocities of the planet-hosting
F5--G5 stars, and of the solar-neighborhood samples.  As expected,
we find that the main-sequence F5--G5 planet-hosting stars are
kinematically hotter than the solar-neighborhood A5--F0 stars, and
they are kinematically indistinguishable from the solar-neighborhood
main-sequence F5--G5 sample.

Figures~\ref{fig02},~\ref{fig03}, and~\ref{fig04} indicate that on
average the subgiant planet-hosting stars are similar in mass to the
main-sequence F5--G5 planet hosts. However it is possible that some
fraction of the subgiant planet hosts stars are indeed more massive than
the F5--G5 planet hosts. To quantify the maximum fraction of massive
stars that could be present in the subgiant sample, we use a Monte Carlo
simulation.  We create many random mixed control samples composed of
both solar neighborhood main-sequence A5--F0 and F5--G5 stars.  We vary
the fraction of main-sequence A5--F0 stars and then compare the velocity
dispersion of the subgiant sample to the resultant mixed control samples.
We find that control samples including less than $\approx\!\!40\%$ main
sequence A5--F0 stars are generally consistent with the subgiant sample.
Consequently, we expect that no more than 40\% of the subgiant planet
hosts were once A5--F0 stars.\footnote{Based on similar tests we also
expect that no more than 60\% of the subgiant planet hosts were once
F0--F5 stars.}  Based on a similar calculation, we also find that no
more than 40\% of the stars in the giant sample were once A5--F0 stars.

We perform a few additional tests of the claim that the subgiants
that were selected for the Doppler planet surveys have a typical mass
similar to that of F5--G5 stars.  First, we repeated the comparisons after
swapping the subgiant planet-hosting sample for the sample of southern sky
subgiants from the Pan-Pacific Planet Search presented by \citet{wit11}.
These are not necessarily planet-hosting stars, but they were selected
for Doppler surveillance in a similar fashion as the other subgiant
planet hosts.  We obtain quantitatively similar results; the Pan-Pacific
search targets are kinematically better matched to F5--G5 stars than to
more massive stars. Second, we select subgiant stars from the Hipparcos
catalog using the same color and magnitude criteria that we applied to
select the planet-hosting subgiant sample.  This is intended to mimic
the parent population of subgiants from which the planet-hosting stars
were identified by the Doppler surveyors.  We find that their kinematics
are indistinguishable from the kinematics of the planet-hosting subgiants
(Anderson-Darling $p$-value $=0.30$).  We also find the Hipparcos subgiant
sample to be kinematically incompatible with an origin in the same parent
population as the Hipparcos main sequence A5--F0 sample (Anderson-Darling
$p$-value $<10^{-6}$).  In contrast, the subgiant planet hosts and the
main-sequence F5-G5 planet hosts are kinematically consistent with an
origin in the same parent population (Anderson-Darling $p$-value $=0.5$).

\subsection{Metallicities}\label{sec:metallicities}

The masses of the subgiant planet-hosting stars are now seen to be
similar to those of the main-sequence F5--G5 planet-hosting stars.
One may wonder, though, if there are systematic differences in other
stellar parameters that may have affected their planet populations.
A key parameter is metallicity, which is well known to have a major
influence on the properties of giant planets \citep{san04,fis05}.  Thus it
is important to compare the metallicities of the two samples.  Based on
the metallicity determinations reported in the literature, we find the
average metallicity of the main-sequence sample ([Fe/H] $=0.05\pm0.03$)
to overlap with the average metallicity of the subgiant sample ([Fe/H]
$=0.03\pm0.03$).  Their distributions are consistent with being drawn
from the same parent distribution (Anderson-Darling $p$-value $=0.3$).
Thus at face value the metallicity distributions appear to be equivalent,
although in Section 5.3 we discuss the possibility of systematic biases
in the reported metallicity values.

\subsection{Orbital Parameters}\label{sec:orbital}

\begin{figure*}
\plotone{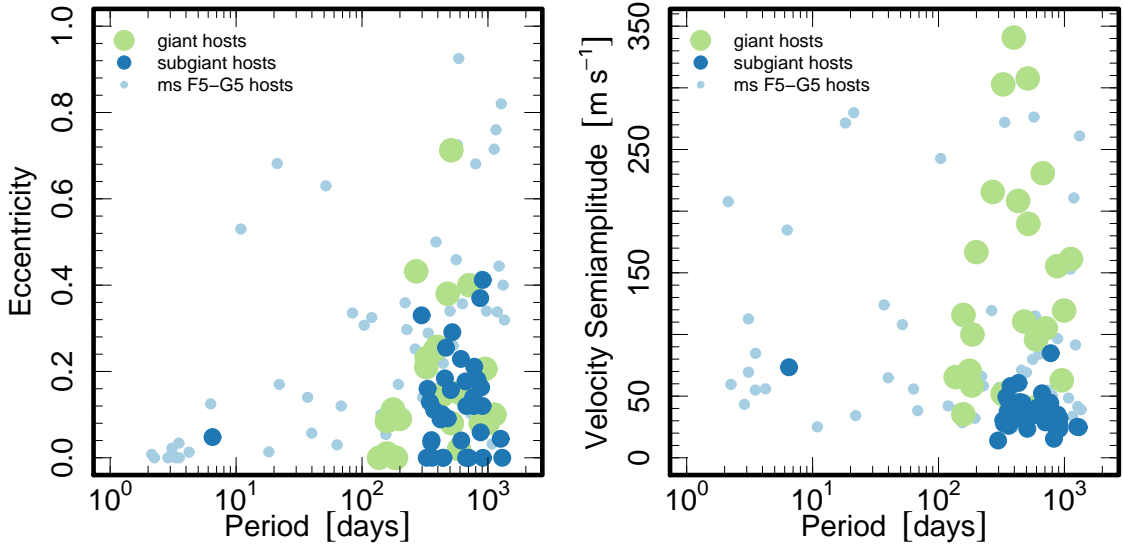}
\caption{Eccentricities, velocity semiamplitudes ($K$), and orbital
periods for the planets in each of our samples: main-sequence F5--G5 stars
(light blue), subgiants (dark blue), and giants (green).  For each sample,
we only include planets that reside in parts of parameter space in which
radial velocity surveys are complete: $K>20$ m s$^{-1}$ and $a < 2.5$ AU
\citep[e.g.,][]{joh10b}.  There are three significant differences between
the samples: (1) there are fewer planets orbiting evolved stars at short
periods, (2) the planets around subgiants have lower eccentricities,
and (3) the planets around subgiants have smaller $K$ values than those
observed around main-sequence stars or giants.\label{fig05}}
\end{figure*}

The radial velocity surveys are complete for planets with velocity
semiamplitude $K>20$ m s$^{-1}$ and orbital distance $a < 2.5$~AU,
corresponding to orbital period $P \lesssim 4$~yr \citep{joh10b}.
We therefore focus attention on the population of known planets satisfying
these criteria.  In Figure~\ref{fig05} we plot the $P$--$e$ and $P$--$K$
distributions for the giant planets with $K>20$ m s$^{-1}$ and $a<
2.5$~AU, for all three samples of planet-hosting stars.

The planets orbiting both samples of evolved stars (subgiants and giants)
mainly have periods longer than 100 days.  Only one subgiant has a
planet with a shorter period, in strong contrast to the main-sequence
stars which frequently host short-period planets.  Furthermore the
planets around the evolved stars are preferentially on circular orbits,
compared to the more eccentric orbits commonly seen for the planets around
main-sequence F5--G5 stars.  The mean eccentricities of the planets around
subgiants, giants, and main-sequence stars are $\bar{e}=0.12\pm0.02$,
$\bar{e}=0.17\pm0.03$, and $\bar{e}=0.36\pm0.04$, respectively.
The eccentricity distributions for planets around subgiants, and for
planets around giants, are consistent with each other (Anderson-Darling
$p$-value $=0.3$).  However the probability that the eccentricities of
the planets around subgiants are drawn from the same parent distribution
as the main-sequence sample is less than $10^{-6}$.  In the next section
we will examine the hypothesis that tidal evolution is responsible for
these differences in the planet populations.

The right panel of Figure~\ref{fig05} also shows that the planets orbiting
subgiants are clustered at low velocity semiamplitudes $K \approx 40$
m s$^{-1}$, relative to the planets around main-sequence stars.  This is
despite the fact that planets with larger $K$ values would have been more
easily detected.  The probability that the $K$ distributions of planets
around subgiants and main-sequence stars are drawn from the same parent
distribution is less than $10^{-6}$.

One finding that is not illustrated in Figure~\ref{fig05} is that
the evolved stars are more likely than the main-sequence stars to
show evidence for additional longer-period companions, in the form
of a long-term radial-velocity trend.  At $P>100$~days, long-term
radial-velocity trends occur around 9 of 34 subgiants.  In contrast,
long-term trends occur in only 2 of 33 main-sequence stars (with giant
planets in the complete region of parameter space) and in only 1 out of
24 giant stars.  If the distribution of long-term trends were the same
for subgiants as for main-sequence stars, the chance of reproducing
these numbers would be less than 1 in 40,000.

\section{Tidal timescales}

If the subgiant planet-hosting stars and the main-sequence F5--G5
planet-hosting stars have similar kinematics and metallicities, the
differences between their planet populations are not easily attributed to
differences in mass or composition.  Instead, the observed differences
in planet populations are likely to be related to age or stellar radius
(which are, of course, correlated).  A mechanism for altering planetary
systems that is known to depend sensitively on stellar radius is tidal
dissipation.  Theoretical rates of energy dissipation due to star-planet
tidal interactions scale as high powers of the stellar radius ($\propto
R_{\ast}^5$ in the works cited below).  In this section we develop
approximate formulas for tidal dissipation rates, to support the
discussion in the next section.

Tidal interactions between a star and an orbiting planet allow for the
exchange of angular momentum between the bodies while steadily draining
energy from the system \citep{cou73,hut80,hut81}.  If the star's
rotation period is longer than the planet's orbital period, the star
spins up and the orbit shrinks.  If the total angular momentum of the
system exceeds a critical value, the orbit eventually circularizes and
the bodies' spins are synchronized and aligned.  If however the total
angular momentum is too small, then there is no stable equilibrium:
the orbit keeps shrinking until the planet is destroyed.  It is not yet
possible to compute accurate dissipation timescales from first principles.
The best that can be done is to establish plausible scaling relationships
that are calibrated using observations, as we attempt here.

Eccentricity damping (orbit circularization) can occur either due to
dissipation inside of the planet or inside of the star.  The timescales
for eccentricity damping due to dissipation inside of the planet
$t_{e,p}$, due to dissipation inside of the star $t_{e,\ast}$, and their
ratio are \citep[e.g.,][]{mar07,maz08}
\begin{eqnarray}
\label{eq:tau-planet}
t_{e,p} & \approx & \frac{2 P}{21} \left(\frac{Q_{p}}{k_{p}}\right) \left(\frac{M_{p}}{M_{\ast}}\right) \left(\frac{a}{R_{p}}\right)^5, \\
\label{eq:tau-star}
t_{e,\ast} & \approx & \frac{2 P}{21} \left(\frac{Q_{\ast}}{k_{\ast}}\right) \left(\frac{M_{\ast}}{M_{p}}\right) \left(\frac{a}{R_{\ast}}\right)^5, \\
\label{eq:ratio-tau}
\frac{t_{e,\ast}}{t_{e,p}} & = & \left(\frac{Q_{\ast}/k_{\ast}}{Q_{p}/k_{p}}\right) \left(\frac{M_{\ast}}{M_{p}}\right)^2 \left(\frac{R_{p}}{R_{\ast}}\right)^5,
\end{eqnarray}

\noindent
where $P$ is the orbital period, $a$ is the orbital semimajor axis,
$Q_{p}$ and $Q_{\ast}$ are the tidal quality factors of the planet and
star, $k_{p}$ and $k_{\ast}$ are their tidal Love numbers, $M_{p}$ and
$M_{\ast}$ are their masses, and $R_{p}$ and $R_{\ast}$ are their radii.
Assuming constant $Q_{p}$, $k_{p}$, $M_{p}$, $R_{p}$, and $M_{\ast}$,
the scaling relations for the timescale for orbit circularization due
to dissipation inside of the planet and inside of its host star are
\begin{eqnarray}
\label{eq:ratio-planet}
\frac{t_{e,p,2}}{t_{e,p,1}} & = & \left(\frac{P_{2}}{P_{1}}\right) \left(\frac{a_{2}}{a_{1}}\right)^5, \\
\label{eq:ratio-star}
\frac{t_{e,\ast,2}}{t_{e,\ast,1}} & = & \left(\frac{P_{2}}{P_{1}}\right) \left(\frac{Q_{\ast,2}/k_{\ast,2}}{Q_{\ast,1}/k_{\ast,1}}\right) \left(\frac{R_{\ast,1}}{R_{\ast,2}}\right)^5 \left(\frac{a_{2}}{a_{1}}\right)^5.
\end{eqnarray}

It is commonly understood that for main-sequence stars with planets,
orbital circularization is mainly due to dissipation within the
planet.  This is because the usual assumptions $Q_{\ast}/k_{\ast}
\sim 10^{6}$, $Q_{p}/k_{p} \sim 10^{5}$, $M_{\ast} = 1~M_{\odot}$,
$M_{p} = 1~M_{\mathrm{Jup}}$, $R_{\ast} = 1~R_{\odot}$, $R_{p} =
1~R_{\mathrm{Jup}}$ lead to $t_{e,\ast}/t_{e,p} \sim 100$.  We use this
fact to provide a rough calibration for the dissipation timescale.
Figure~\ref{fig05} shows that planets orbiting main-sequence stars
inside of $P=5$ days are on nearly circular orbits, while planets
orbiting outside of $P=10$ days are frequently on eccentric orbits.
The timescale for orbit circularization in the shorter-period systems
must be significantly shorter than the ages of the systems.  We will
therefore assume that the circularization timescale $\tau_{e,p}$ at $P =
5$ days for a solar-mass star with $Q_{\ast}/k_{\ast} \sim 10^{6}$ and a
planet with the mass and radius of Jupiter and $Q_{p}/k_{p} \sim 10^{5}$
is $\tau_{e,p} = 1$ Gyr.

When the host star is a subgiant with $R_{\ast} = 4~R_{\odot}$,
then $t_{e,\ast}/t_{e,p} \sim 0.01$ and circularization should be
mainly due to dissipation inside of the star \citep[e.g.,][]{ras96}.
Using Equations~(\ref{eq:ratio-tau}), (\ref{eq:ratio-planet}), and
(\ref{eq:ratio-star}) plus the calibration presented above we may write
\begin{eqnarray}
t_{e,\ast} & = & \tau_{e,p} \Theta \left(\frac{P}{\mathrm{5~days}}\right) \left(\frac{a}{\mathrm{0.06~AU}}\right)^5,
\end{eqnarray}

\noindent
where
\begin{eqnarray}
\Theta & \equiv & \left(\frac{Q_{\ast}/k_{\ast}}{Q_{p}/k_{p}}\right) \left(\frac{M_{\ast}}{M_{p}}\right)^2 \left(\frac{R_{p}}{R_{\ast}}\right)^5.
\end{eqnarray}

\noindent
Then, taking the overall eccentricity-damping timescale $t_e$ to be the
smaller of the two dissipation timescales (inside the planet or 
inside the star), we have
\begin{eqnarray}
\label{eq:tau-circ}
t_{e} & = & \tau_{e,p} \left(\frac{P}{\mathrm{5~days}}\right) \left(\frac{a}{\mathrm{0.06~AU}}\right)^5 \min{(1,\Theta)}.
\end{eqnarray}

As for tidal evolution of the orbital distance $a$, the scaling relation
is thought to be \citep[e.g.,][]{lin96}
\begin{eqnarray}
t_{a} & \propto & \frac{M_{\ast}^{1/2} (Q_{\ast}/k_{\ast}) a^{13/2}}{M_{p} R_{\ast}^5}.
\end{eqnarray}

\noindent
Figure~\ref{fig05} shows that there are many planets orbiting main
sequence stars at $P \approx 5$ days, but few at $P \approx 1$
day. Assuming this difference to reflect the tidal decay of the
shortest-period orbits leads to a rough calibration of the tidal
decay timescales. If we assume that the timescale for orbital drift
$\tau_{a}$ of a $1~M_{\mathrm{Jup}}$ planet orbiting at $P = 5$ days
a $Q_{\ast}/k_{\ast} \sim 10^{6}$, $1~M_{\odot}$, and $1~R_{\odot}$
host star is $\tau_{a} = 10$ Gyr, then we have
\begin{eqnarray}
\label{eq:tau-a}
t_{a} & = & \tau_{a} \left(\frac{Q_{\ast}/k_{\ast}}{10^{6}}\right) \left(\frac{R_{\ast}}{R_{\odot}}\right)^{-5} \left(\frac{a}{\mathrm{0.06~AU}}\right)^{13/2}.
\end{eqnarray}

\begin{figure*}
\plotone{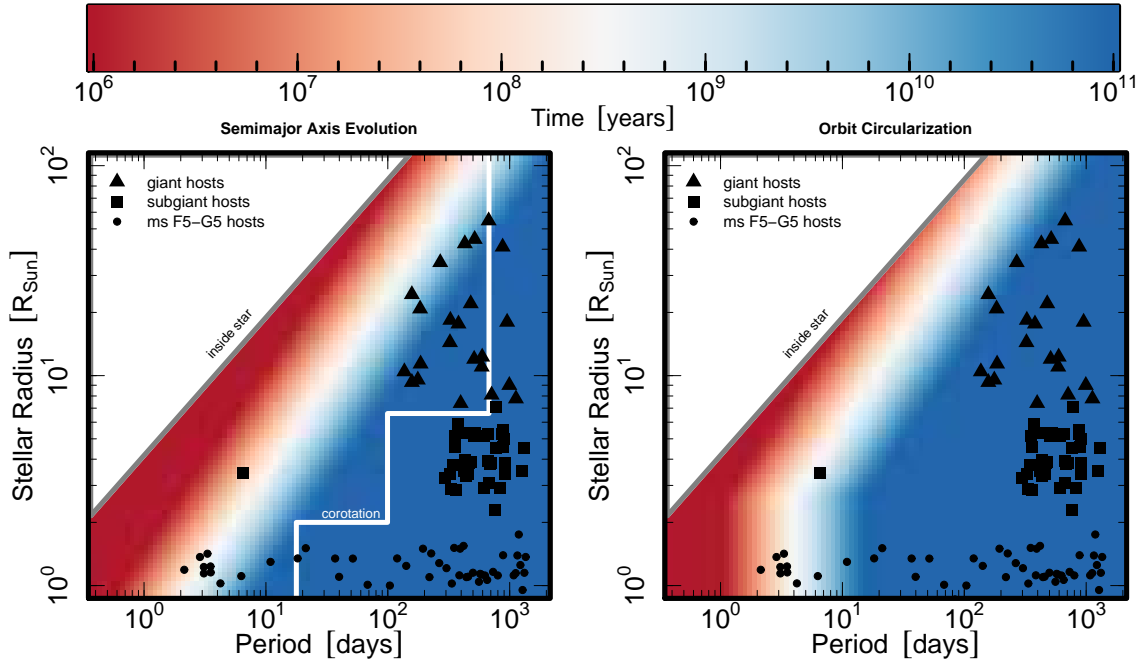}
\caption{Tidal timescales as a function of period and stellar radius.
{\it Left}: Timescale $t_{a}$ for semimajor axis evolution (orbital decay)
for planets inside the corotation radius (marked approximately by the
white line).  {\it Right}: Timescale $t_{e}$ for orbit circularization.
The timescales were computed as described in the text, assuming
$M_{\ast} = 1~M_{\odot}$, $M_{p} = 1~M_{\mathrm{Jup}}$, $R_{p} =
1~R_{\mathrm{Jup}}$, $Q_{\ast}/k_{\ast} = 10^{6}$, and $Q_{p}/k_{p}
= 10^{5}$.  For planets to the left of the white diagonal band, the
calculated tidal timescale is short compared to the duration over which
a solar-mass, solar-metallicity star exists as a subgiant or at the base
of the giant branch \citep[e.g.,][]{dot08}.  The left-hand figure shows
that the scarcity of $P<10$ day planets around subgiants is a seemingly
inevitable consequence of tidal decay.  The right-hand figure shows that
the tendency for planets around subgiants to have nearly circular orbits
all the way out to $P \approx 200$ days can only be explained if stars
become much more dissipative as they evolve off of the main sequence
($Q_{\ast}/k_{\ast} \sim 10^2$).  Such a scenario would also imply
enhanced rates for tidal decay, and could explain the scarcity of planets
around subgiants all the way out to $P \approx 100$ days.\label{fig06}}
\end{figure*}

Equations (\ref{eq:tau-circ}) and (\ref{eq:tau-a}) were used to calculate
$t_e$ and $t_a$ as a function of orbital period and stellar radius,
for a fiducial solar-mass star with a Jovian planet. The results are
shown in Figure~\ref{fig06}, which also displays the periods and stellar
radii of the various samples of exoplanet systems considered in this
paper. The stellar radii are based on the empirical calibration given
by \citet{tor10}.

There are two other criteria that must be satisfied for tidal decay.
First, since tides only cause orbital decay if the stellar rotation period
is longer than the orbital period, we indicate in Figure~\ref{fig06} the
maximum stellar rotation period that would be expected for stars in the
different samples.  These are based on the observed projected rotation
rates ($v\sin{i}$) and estimated radii of the planet host stars, using
\begin{eqnarray}
P_{\mathrm{rot}} & \lesssim \frac{2 \pi R_{\ast}}{v\sin{i}}.
\end{eqnarray}

\noindent
The upper limit on $P_{\rm rot}$ was computed for each star. Then, the
median of these upper limits was computed for each sample, and plotted
in the left panel of Figure~\ref{fig06}.

Second, the total angular momentum of the system $L_{\mathrm{tot}}$
must be lower than the critical value $L_{\mathrm{crit}}$ necessary for a
stable equilibrium (in which the star and planet have achieved spin--orbit
synchronization and alignment). The total angular momentum can be written
\begin{eqnarray}
L_{\mathrm{tot}} & = & L_{\mathrm{orb}} + L_{\mathrm{rot},\ast} + L_{\mathrm{rot},p}, \\
L_{\mathrm{orb}} & = & \mu \left[G M_{\mathrm{tot}} a (1-e^2) \right]^{1/2}, \\
L_{\mathrm{rot},\ast} & = & g_{\ast}^2 M_{\ast} R_{\ast}^2 \omega_{\ast}, \\
L_{\mathrm{rot},p} & = & g_{p}^2 M_{\ast} R_{p}^2 \omega_{p},
\end{eqnarray}

\noindent
where $\mu = (M_{\ast} M_{p})/(M_{\ast}+M_{p})$ is the reduced mass and
$M_{\mathrm{tot}} = M_{\ast}+M_{p}$ is the total mass of the system. Here
$g_{\ast}^2 = 0.06$ and $g_{p}^2 = 0.25$ are the radii of gyration of star
and planet, assuming the solar and Jovian values respectively.  Likewise,
$\omega_{\ast}$ and $\omega_{p}$ are the angular velocities of star and
planet.  The rotational angular momentum of the planet is negligible.
The value of $L_{\mathrm{crit}}$ is \citep[e.g.,][]{hut80,mat10}
\begin{eqnarray}
L_{\mathrm{crit}} & = & 4 \left[\frac{G^2}{27} \frac{M_{\ast}^3 M_{p}^3}{M_{\ast}+M_{p}} \left(I_{\ast} + I_{p}\right)\right]^{1/4},
\end{eqnarray}

\noindent
where $I_{\ast} = g_{\ast}^2 M_{\ast} R_{\ast}^2$ and $I_{p}
= g_{p}^2 M_{p} R_{p}^2$.  Assuming $M_{\ast} = 1~M_{\odot}$,
$M_{p} = 1~M_{\mathrm{Jup}}$, $R_{\ast} = 4~R_{\odot}$, and $R_{p} =
1~R_{\mathrm{Jup}}$ characteristic of an evolved star with a giant planet,
then all systems with $a \lesssim 0.35 \mathrm{~AU}$, or approximately
$P \lesssim 75 \mathrm{~days}$, are tidally unstable.  Such planets will
ultimately be engulfed, if the host star's rotational angular momentum
was initially comparable to the current solar value.

\section{Discussion}

\subsection{Hot Jupiters}

The left panel of Figure~\ref{fig06} shows that there is no difficulty
explaining the lack of hot Jupiters around subgiants with periods between
1 and 10 days.  Although such planets' orbits may be stable for a Hubble
time around a main-sequence stars, once the host star evolves and reaches
a radius of $4~R_{\odot}$, tidal dissipation rates increase by a factor
of $\sim\!\!10^2$.  Those same planets would experience orbital decay
in a few 100 Myr, comparable to the time that their host stars spend as
subgiants \citep[e.g.,][]{dot08}.  Tidal destruction has already been
noted as a theoretical possibility by \citet{vil09} and \citet{kun11},
at least for giant stars with $R_{\ast} \gtrsim 20~R_{\odot}$.  It has
also been suggested by \citet{llo11}, among others.  Now that the
subgiants are seen to have similar masses as the main-sequence sample,
tidal destruction seems to be the inevitable explanation for the lack of
close-in giant planets orbiting subgiants relative to main-sequence stars.

One might wonder if it is possible to find additional supporting evidence
for tidal destruction of hot Jupiters based on the observed rotation rates
of subgiants, since the accretion of angular momentum from a disrupted
giant planet could affect the rotation rate.  Suppose a star with mass
$1.3~M_{\odot}$, radius $1.1~R_{\odot}$, radius of gyration $I/MR^2 =
0.06$ (the solar value), and rotation period $P_{\mathrm{rot}} = 10$
d accretes a 1 $M_{\rm Jup}$ planet initially orbiting at 0.1 AU,
and absorbs all of its orbital angular momentum. The final rotation
period of that star when it reaches $R_{\ast} = 4~R_{\odot}$ would be 24
days. Had it not absorbed its planet's orbital angular momentum, then its
rotational period would have been 130 days.  This seems like a significant
difference.  However, the radii of subgiants are generally uncertain by a
factor of two or more. If instead the subgiant had a radius of $R_{\ast}
= 8~R_{\odot}$, then its rotation period after absorbing its planet's
orbital angular momentum would be 94 days.  Consequently, identifying
angular momentum enhancement in evolved stars will be difficult without
accurate stellar radii.

The accretion of metal-enriched material from a disrupted giant planet
could also affect the observed chemical abundances of an evolved star.
Indeed, the accretion of solid material was initially considered as one
possible explanation for the enhanced metallicity of the solar-type stars
identified as giant planet hosts \citep[e.g.,][]{lau00}.  Following the
calculation by \citet{lau00}, the accretion of a Jupiter-mass planet
with $Z \approx 0.1$ into the $0.5~M_{\odot}$ convective envelope of an
evolved solar metallicity star will only increase the star's observed
metallicity by 0.01 dex -- an imperceptibly small increase.

\subsection{Intermediate-period Giant Planets}

Although tidal destruction is a natural explanation for the lack of hot
Jupiters around subgiants, the relative scarcity of giant planets with
periods between 10 and 100 days cannot be explained as the consequence
of tidal destruction unless the tidal dissipation rate for subgiants
is 3-4 orders of magnitude faster than estimated in the previous
section (and plotted in Figure~\ref{fig06}).  One possibility is that
dissipation becomes faster not only because of an enlarged stellar
radius, but also a decrease in $Q_{\ast}/k_{\ast}$ due to the changing
stellar structure.  This could also explain the observation that the
shortest-period planets seen around subgiants ($P \approx 200$ days)
tend to have circular orbits: such planets would be undergoing tidal
eccentricity damping due to dissipation within the host star.  This is
in contrast to the hot Jupiters observed around main-sequence stars, for
which dissipation within the planet is likely to be the primary cause
of eccentricity damping.  Furthermore it would not be surprising for a
subgiant to have a lower $Q_{\ast}/k_{\ast}$ than a main-sequence star.
The strength of tidal evolution in evolved stars is thought to depend
on the mass of their convective envelopes \citep{ras96}, and the mass
in the convective envelope of a solar-type star increases by an order
of magnitude as a subgiant between the main sequence turnoff and the
base of the red giant branch \citep{sac93}.

We digress momentarily to point out that Figure~\ref{fig01} reveals that
most of the {\it giant} planet-hosting stars (as opposed to subgiants)
are concentrated in the so-called ``red clump'' --- a region of the HR
diagram where the red giant branch overlaps with the horizontal branch
for a solar-metallicity stellar population.  Without asteroseismic
observations, we cannot confidently assign individual stars to one or the
other of those two populations.  Unfortunately this means the giants are
less useful than the subgiants in understanding the tidal evolution of
planetary systems, for the following reason.  The stars on the horizontal
branch are currently burning helium in their cores after an initial
ascent up the giant branch.  They very likely had larger radii at the
tip of the red giant branch than they currently possess.  Since tidal
effects depend strongly on stellar radius, the tidal evolution of any
close-in exoplanets depends strongly on the sizes of the giant stars
at the tip of the red giant branch --- which are unobserved and poorly
constrained. The subgiants invite a more straightforward interpretation,
since they are presently as large as they have ever been since the
planet-formation epoch.

Putting these pieces together, a scenario that explains many of
our observations is as follows.  Stars become more dissipative as
they evolve off of the main sequence, both because the stellar radius
increases and also because $Q_{\ast}/k_{\ast}$ decreases by 3-4 orders of
magnitude. Consequently, planets orbiting inside of the corotation radius
($P\lesssim 100$ days) lose angular momentum relatively quickly through
tidal interactions.  Planets just outside of corotation are pushed out
and pile up at $P \approx 200$ days, while also undergoing eccentricity
damping.  Extending this scenario to the giant stars is also possible.
As the host stars continue to move up the red giant branch, they expand
and their rotation slows.  As a consequence, the corotation radius is
enlarged.  Those planets that were previously secure at $P \approx 200$
days now find themselves losing orbital angular momentum.  These planets
are not necessarily destroyed, though, because the host stars' trip up
the red giant branch is brief. The orbital distances may shrink only
modestly before the host stars start burning helium in their cores,
contract, and settle onto the horizontal branch.

\subsection{Occurrence Rates of Giant Planets}

Some of the observed differences between the planet populations around
subgiants and main-sequence stars have no obvious interpretation in
terms of tidal interactions.  First, there is no simple explanation
for the higher occurrence rate of long-period giant planets around
subgiant stars. According to \citet{bow10}, the occurrence rate of
giant planets with $a<3$ AU is $26^{+9}_{-8}$\% as compared to 5--10\%
for main-sequence stars.  More recently, \citet{joh10b} accounted for
the effect of metallicity on planet occurrence and determined that the
occurrence rate of giant planets with $a\lesssim2.5$ AU is $11\pm2$\%
for subgiants, as compared to $6.5\pm0.7$\% for main-sequence stars.
In either case, these offsets represent a 2-3$\sigma$ difference.

A second and related point is that tides cannot explain
the higher incidence of velocity trends noted at the end of
Section~\ref{sec:orbital}.  Tides are too weak to affect planets with
such long periods that they are observed only as linear trends in a
subgiant's radial velocity curve.

Third, and most puzzling, is the clustering of $K$ values between 10--50
m~s$^{-1}$ for the subgiants, as compared to the broader distributions
10--300 m~s$^{-1}$ for planets in the same period range around both
main-sequence stars and giants.  The timescale for tidal evolution is
proportional to planet mass, raising the possibility that only low-mass
planets (small $K$ values) persist once a star evolves off of the main
sequence. But any tidal explanation would struggle to explain why the
giant stars (which are even more evolved than subgiants) still harbor
more massive planets.  Future work is needed to understand these three
differences.

One possibility is that the average metallicity of the stars in the
subgiant sample is higher than that of the main-sequence sample,
despite the comparison made in Section~\ref{sec:metallicities}
suggesting that the metallicities are similar.  Giant planet occurrence
rates are known to be strongly dependent on host star metallicity
\citep[e.g.,][]{san04,fis05}. In particular \citet{fis05} found that
the probability of giant planet occurrence $P_{p}$ scales as
\begin{eqnarray}
P_{p} \propto 10^{2\left[\textrm{Fe/H}\right]}.
\end{eqnarray}

\noindent

In that case, the increased planet occurrence rate for subgiants
reported by \citet{joh10b} could be explained by a residual systematic
metallicity offset of $\log{(\sqrt{11/6.5})} \approx 0.1$ dex in [Fe/H].
At the $1\sigma$ limits of the \citet{joh10b} occurrence rates, even a
residual offset as small as $\log{(\sqrt{9/7.2})} \approx 0.05$ dex in
[Fe/H] could explain the apparently larger occurrence rate of giant
planets in the subgiant sample.

Such an offset does not seem out of the question. The metallicities of the
subgiants and many FG planet host stars in the control sample have been
measured with the Spectroscopy Made Easy package \citep[SME;][]{val96} and
other similar spectral-fitting codes. The metallicity values returned by
SME are known to have systematic biases depending on effective temperature
$T_{\mathrm{eff}}$ \citep[e.g., Section 6.4 of][]{val05}. Although
\citet{val05} attempt to correct for this bias, perhaps there still exists
a residual offset of 0.1~dex in [Fe/H] when comparing the main-sequence
F5--G5 sample ($T_{\mathrm{eff}} \approx 6000$~K) to the subgiant sample
($\approx$5000~K). Indeed, strong correlations between the values of
$T_{\mathrm{eff}}$ and [Fe/H] generated by SME were recently identified
by \citet{tor12}, who demonstrated that SME reports cooler temperatures
(100~K) and higher metallicities (0.1~dex) than a line-by-line MOOG-like
analysis \citep{sne73}.  A homogeneous MOOG-like line-by-line stellar
parameter analysis of a large sample of both FGK dwarf and subgiant
planet-host stars might help to resolve this issue.

We also note that in calculating the dependence of giant planet occurrence
rates on stellar mass and metallicity, \citet{joh10b} assumed that the
effects of mass and metallicity were separable, even though there is
a significant positive correlation between the reported stellar masses
and metallicities (see their Figure 2). As we have shown, the subgiant
stars are not likely to be especially massive. As a result, some of the
increased planet occurrence signal in the subgiant sample was likely
misattributed to mass instead of metallicity.  For that reason, the
metallicity corrections necessary to directly compare the giant planet
incidence rate of main-sequence FGK dwarfs and subgiants may have been
underestimated by \citet{joh10b}.

More speculative is the possibility raised by \citet{llo11} that some of
the radial-velocity signals detected for subgiant stars do not represent
planets, but rather stellar oscillations.  The oscillations would
need to be nonradial to avoid producing large brightness variations
in contradiction with the observations.  Nonradial oscillations
can produce low-amplitude, sinusoidal radial velocity variations
\citep[e.g.,][]{hat96}, though usually with smaller amplitudes and shorter
periods.  To our knowledge nonradial oscillations of subgiants producing
signals as large as tens of m s$^{-1}$ with periods of hundreds of days
have never been empirically established, nor have they been theoretically
predicted.  We note, though, that the literature does contain examples
of similar quasi-sinusoidal variations that are at least suspected to
be the result of nonradial oscillations, in both more evolved and less
evolved stars \citep{hat99,des09}.  Perhaps the intense scrutiny of the
subgiants by the Doppler surveyors has revealed a new mode of stellar
pulsation.  In addition to bringing the occurrence rate of giant planets
around subgiants into better agreement with the main-sequence stars,
this might help to explain the clustering of $K$ values.  The physics of
the hypothetical pulsations might naturally select a particular range
of velocities, whereas the planet-induced velocity amplitudes would
depend on three different independent variables ($M_{p}$, $i$, and $P$)
and should combine to produce a wide range of $K$ values.

\subsection{Stellar Mass Loss}

One might also wonder if stellar mass loss contributes to the orbital
evolution of planets around subgiant stars, in addition to tidal effects.
Mass loss on the red giant branch is frequently parametrized using the
Reimers relation \citep[e.g.,][]{kud78}
\begin{eqnarray}
\frac{d M_{\ast}}{dt} & = & \eta \left(4\times10^{-13}\right) \frac{L_{\ast}(t) R_{\ast}(t)}{M_{\ast}} \frac{M_{\odot}}{\mathrm{yr}},
\end{eqnarray}

\noindent
where $L_{\ast}$, $R_{\ast}$, and $M_{\ast}$ are the stellar luminosity,
radius, and mass in solar units.  The dimensionless parameter $\eta$ is
in the range $0.4 \lesssim \eta \lesssim 0.8$ \citep[e.g.,][]{sac93} and
frequently set to $\eta = 0.5$ \citep[e.g.,][]{kud78,hur00}.  Taking the
median luminosity, radius, and isochrone mass of our subgiant planet
host sample $L_{\ast} = 9.5~L_{\odot}$, $R_{\ast} = 3.8~R_{\odot}$,
and $M_{\ast} = 1.6~M_{\odot}$, the expected mass loss is rate is $d
M_{\ast}/dt \approx 5 \times 10^{-12}~M_{\odot}$~yr$^{-1}$. Over the
500 Myr a solar-type star spends as a subgiant, the total mass lost
is only $M_{\mathrm{loss}} \lesssim 0.01~M_{\odot}$.  Thus, mass loss
is unlikely to affect the orbital evolution of close-in giant planets
orbiting subgiants or stars at the base of the red giant branch.

\section{Conclusions}

Subgiant stars are observed to host fewer close-in giant planets than
main-sequence stars, but have a systematically higher giant planet
occurrence rate when integrated over all periods.  These differences
have been attributed to the increased mass of the subgiant planet
hosts relative to main-sequence planet hosts.  We find that the
Galactic kinematics of subgiant stars that host giant planets demand
that on average they be similar in mass to solar-type main-sequence
planet-hosting stars.  Therefore, the best explanation for the lack
of hot Jupiters around evolved stars is tidal destruction of the hot
Jupiters they once possessed.  If tidal dissipation inside their host
stars is also responsible for circularizing the orbits of the giant
planets orbiting subgiants at $P \approx 200$ days, then tides should
also have been strong enough to destroy any giant planets inside of
corotation ($P \lesssim 100$ days).  Planets outside of corotation
may have extracted angular momentum from their host stars and moved
to longer orbital periods, explaining the sudden appearance of planets
around subgiant stars at $P \approx 200$ days.  The apparent 2-3$\sigma$
increase in the incidence of long-period giant planets around subgiant
stars relative to main-sequence stars is difficult to explain.  It may be
the result of systematic underestimates of the metallicities of subgiant
stars or (more speculatively) hitherto unknown modes of stellar pulsation
among the subgiants that masquerade as long-period giant planets.

It is also important to note that the kinematic equivalence of the
subgiant and main-sequence samples suggests that the theoretical
stellar-evolutionary models that were used to infer larger masses for
the subgiants need to be re-evaluated. In particular, as suggested by
\citet{llo11} it may be necessary to reassess the applicability of
a mixing-length model of convection that is calibrated for the Sun
to low surface gravity stars that are not even quasi-stationary
and that are developing shells of superadiabatic convection
\citep[e.g.,][]{rob04,cas12}.

There are also potentially important implications of our suggestion
that stars become much more dissipative as they evolve off of the main
sequence, such that $Q_{\ast}/{k_\ast} \sim 10^{2-3}$.  If this is
so, then the Earth will very likely be destroyed as the Sun expands
to nearly 100 times its current size near the tip of the red giant
branch.  Previously it was thought that the orbital evolution of
the Earth could not even be predicted qualitatively as it seemed
to depend on the details of the Sun's mass loss on the giant branch
\citep[e.g.,][]{sac93,ras96}.  Thus the subgiant planet searches may
have---quite unexpectedly---clarified the Earth's ultimate fate.

\acknowledgments We thank John Johnson, Adam Kraus, Greg Laughlin, Jamie
Lloyd, Nevin Weinberg, and the anonymous referee for their comments and
constructive criticism. This research has made use of NASA's Astrophysics
Data System Bibliographic Services, the Exoplanet Orbit Database and the
Exoplanet Data Explorer at exoplanets.org, and both the SIMBAD database
and VizieR catalogue access tool, CDS, Strasbourg, France.  The original
description of the VizieR service was published by \citet{och00}.
It has also used The IDL Astronomy User's Library, a service of the
Astrophysics Science Division (ASD) at NASA's GSFC. Support for this
work was provided by the MIT Kavli Institute for Astrophysics and Space
Research through a Kavli Postdoctoral Fellowship.

\end{document}